\documentclass[aps,prd,onecolumn,12pt,tightenlines,groupedaddress,longbibliography]{revtex4-2}

\usepackage{amsmath,amssymb,amsfonts}
\usepackage[T1]{fontenc}
\usepackage{graphicx}
\usepackage{array}
\usepackage{adjustbox}
\usepackage{multirow}
\usepackage{braket}
\usepackage{hyperref}
\usepackage{bm}

\begin{document}

\title{New Pairing Mechanism via Chiral Electron-Hole Condensation for Non-BCS Superconductivity}


\author{Wanpeng Tan}
\email[]{wtan@nd.edu}
\affiliation{Department of Physics and Astronomy, University of Notre Dame, Notre Dame, Indiana 46556, USA}

\date{\today}

\begin{abstract}
A novel chiral electron-hole (CEH) pairing mechanism is proposed to account for non-BCS superconductivity. In contrast to BCS Cooper pairs, CEH pairs exhibit a pronounced affinity to antiferromagnetism for superconductivity. The gap equations derived from this new microscopic mechanism are analyzed for both s- and d-wave superconductivity, revealing marked departures from the BCS theory. Unsurprisingly, CEH naturally describes superconductivity in strongly-correlated systems, necessitating an exceedingly large coupling parameter ($\lambda>1$ for s-wave and $\lambda>\pi/2$ for d-wave) to be efficacious. The new mechanism provides a better understanding of various non-BCS features, especially in cuprate and iron-based superconductors. In particular, CEH, through quantitative comparison with experimental data, shows promise in solving long-standing puzzles such as the unexpectedly large gap-to-critical-temperature ratio $\Delta_0/T_c$, the lack of gap closure at $T_c$, superconducting phase diagrams, and a non-zero heat-capacity-to-temperature ratio $C/T$ at $T=0$ (i.e., the ``anomalous linear term''), along with its quadratic behavior near $T=0$ for d-wave cuprates.
\end{abstract}


\maketitle


\section{\label{sec_intro}Introduction}

Magnetism has traditionally been viewed as antagonistic to conventional Bardeen-Cooper-Schrieffer (BCS) superconductivity \cite{bardeen1957}. However, many non-BCS superconductors discovered in the past few decades have demonstrated the opposite, showing that strong magnetism is actually very conducive to non-BCS superconductivity. In particular, many of them are derived from parent compounds with antiferromagnetic properties, and some even exhibit cases of the coexistence of superconductivity and antiferromagnetic order \cite{bennemann2008}. The two primary classes of high $T_c$ superconductors, cuprate \cite{bednorz1986} and iron-based (FeSC) \cite{kamihara2006} superconductors, both have their roots in antiferromagnetic compounds, with cuprates originating from antiferromagnetic Mott insulators and FeSCs from antiferromagnetic metals.

The coexistence of superconductivity and long-range antiferromagnetic order was actually discovered a long time ago in the late 1970s \cite{ishikawa1977,mccallum1977}. The intimate relationship between antiferromagnetism and superconductivity has been observed in many different types of non-BCS superconducting materials, including heavy fermion compounds \cite{steglich1979,mathur1998}, organic superconductors such as quasi-1D TMTTF/TMTSF type and quasi-2D BEDT-TTF type \cite{lebed2008}, and doped fullerenes \cite{takabayashi2016}.

The significance of antiferromagnetic order may provide crucial clues for solving the puzzles of non-BCS superconductivity. In particular, this leads to our proposal of a new pairing mechanism via chiral electron-hole (CEH) condensation in this work. The presence of strong antiferromagnetic correlations is critical, as it guarantees that chirally opposite electron and hole states are next to each other, making the chiral condensation more feasible. This may explain why both high $T_c$ superconducting classes (cuprates and FeSCs) are based on antiferromagnetic compounds.

In order to fully comprehend these superconductors, it is imperative to not only find their correct pairing mechanism, but also to identify their pairing symmetry. Angle-resolved photoemission spectroscopy measurements have indisputably proven that cuprates display a $d_{x^2-y^2}$ gap symmetry \cite{shen1993}. Although studies on the pairing symmetry of FeSCs that were discovered much later are not as conclusive, the majority consensus suggests that they most probably exhibit some type of s-wave pairing symmetry \cite{zhang2012}.

In this paper, we apply the mean-field approach to derive superconducting gap equations using the new pairing mechanism. Detailed analysis of the equations for both s-wave and d-wave superconductivity will reveal various features that differ from BCS. Furthermore, we will address puzzles concerning superconducting gap and heat capacity and present several examples that directly compare CEH predictions with cuprate and FeSC data. Natural units ($\hbar=c=k_B=1$) are utilized throughout the work for convenience.

\section{\label{ceh}Chiral Electron-Hole (CEH) Pairing}

We will closely follow the Bogoliubov-BCS formalism as described in Ref. \cite{tinkham2004} for the mean-field theory of BCS superconductivity, albeit with a new microscopic superconducting mechanism. First of all, we will begin with a four-fermion interacting Hamiltonian for the simple straightforward case of s-wave CEH pairing,
\begin{equation}\label{eq:4fermi}
H=\sum_{\mathbf{k}\sigma}\xi_{\mathbf{k}} c^\dagger_{\mathbf{k}\sigma} c_{\mathbf{k}\sigma}-V\sum_{\mathbf{k}\mathbf{k}'} c^\dagger_{\mathbf{k}L} c_{-\mathbf{k}R} c^\dagger_{-\mathbf{k}'R} c_{\mathbf{k}'L}
\end{equation}
where we use left and right chiralities instead of the conventional up and down spin notation to emphasize the significance of chirality in this study.

Note that similar four-fermion interactions were also used in the Nambu-Jona-Lasinio (NJL) mechanism \cite{nambu1961} in particle physics for quark condensation and spontaneous symmetry breaking, borrowing the idea from the earlier BCS superconductivity work \cite{bardeen1957}. Such ideas are also crucial in the recently developed mirror matter theory, which aims to address many puzzles in fundamental physics and cosmology \cite{tan2019,tan2019c,tan2021a,tan2023a,tan2023b}. In particular, the concept of staged chiral quark condensation \cite{tan2019c,tan2019e,tan2023b} has directly motivated this work.

The most significant difference in Eq. \ref{eq:4fermi} from the BCS Hamiltonian is that, by borrowing back the idea of the NJL model, the four creation and annihilation operators are arranged to incorporate the proposed condensation mechanism of chiral electron-hole pairs instead of the conventional Cooper pairs. Specifically, the superconducting pairs are formed from electrons and holes with exactly opposite chiralities, a configuration readily achievable in adjacent sites of antiferromagnetic materials, which also explains the non-local behavior of the CEH mechanism. We can then define a similar order parameter $\bm{\Delta}$ based on the CEH condensation mechanism,
\begin{eqnarray}\label{eq:delta}
\bm{\Delta} = V\sum_{\mathbf{k}} \braket{c^\dagger_{\mathbf{k}L}c_{-\mathbf{k}R}},\;\;\;\; \bm{\Delta}^*=V\sum_{\mathbf{k}} \braket{c^\dagger_{-\mathbf{k}R}c_{\mathbf{k}L}}.
\end{eqnarray}

\begin{figure}
\includegraphics[scale=1.0]{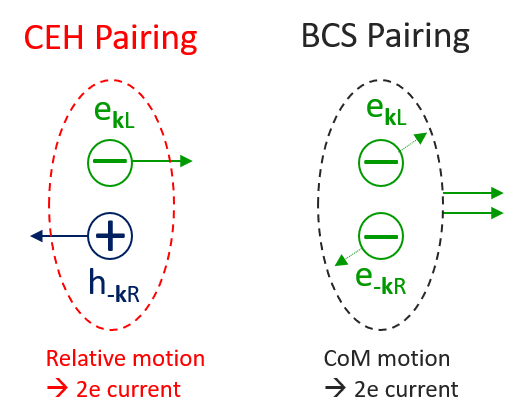}
\caption{\label{fig:pair}Different pairing and conducting mechanisms between CEH and BCS are shown.}
\end{figure}

Upon initial instinctive consideration, one might assume that CEH pairs cannot conduct electric currents due to their zero net charge. However, the conduction mechanism of CEH pairs is fundamentally different from that of Cooper pairs, as illustrated in Fig. \ref{fig:pair}. While the Cooper pair conducts currents through center-of-mass motion, the CEH pair achieves this through relative motion. On a macroscopic scale, both mechanisms yield equivalent $2e$-like currents, resulting in comparable outcomes in most macroscopic phenomena, including the Josephson effect.

Another important aspect concerns the pairing symmetry. In CEH condensation, the pairs must be spin singlets owing to its chiral nature (like the Higgs in particle physics), leading to symmetric orbital wave functions. Consequently, the resulting pairing symmetry can only be s-, d-, or g-wave.

Considering the CEH condensation, the Hamiltonian of Eq. \ref{eq:4fermi} then takes the bilinear form,
\begin{equation}\label{eq:cond}
H=\sum_{\mathbf{k}} \begin{pmatrix}c^\dagger_{\mathbf{k}L},& c^\dagger_{-\mathbf{k}R} \end{pmatrix}
\begin{pmatrix} \xi_{\mathbf{k}} &  -\bm{\Delta}^* \\
-\bm{\Delta} & \xi_{\mathbf{k}} \end{pmatrix}
\begin{pmatrix} c_{\mathbf{k}L}\\ c_{-\mathbf{k}R} \end{pmatrix}.
\end{equation}
We can then diagonalize the Hamiltonian through the Bogoliubov transformation \cite{bogoljubov1958} as follows,
\begin{equation}
U^\dagger \begin{pmatrix} \xi_{\mathbf{k}} &  -\bm{\Delta}^* \\
-\bm{\Delta} & \xi_{\mathbf{k}} \end{pmatrix} U = \begin{pmatrix} E^+_{\mathbf{k}} &  0 \\
0 & E^-_{\mathbf{k}}  \end{pmatrix}, \;\;\;\; U=\begin{pmatrix} u_{\mathbf{k}} &  v^*_{\mathbf{k}} \\
-v_{\mathbf{k}} & u^*_{\mathbf{k}}  \end{pmatrix}
\end{equation}
where the eigenvalues are,
\begin{equation}\label{eq:ene}
E^{\pm}_{\mathbf{k}} = \xi_{\mathbf{k}} \pm |\Delta|,
\end{equation}
in contrast to $E^{\pm}_{\mathbf{k}} = \pm\sqrt{\xi^2_{\mathbf{k}} + |\Delta|^2}$ in BCS. The corresponding emergent Bogoliubov quasi-particles (energy eigenstates) are therefore defined as follows,
\begin{equation}
\begin{pmatrix} b_{\mathbf{k}1}\\ b_{-\mathbf{k}2} \end{pmatrix} = U^\dagger
\begin{pmatrix} c_{\mathbf{k}L}\\ c_{-\mathbf{k}R} \end{pmatrix}
\end{equation}
where the quasi-particle operators $b$ and $b^\dagger$ satisfy the same anticommutation relations as fermions.
Applying the unitarity condition of $|u|^2+|v|^2=1$, we arrive at the solution,
\begin{equation}
|u|=|v|=\frac{1}{\sqrt{2}}
\end{equation}
which is remarkably different from the BCS findings. To facilitate later discussion, we can introduce a phase factor by setting $u/v=e^{i\delta}$, which gives $u^*v=1/2e^{-i\delta}$.

Using the above solution, we obtain the following condensation relation
\begin{equation}\label{eq:cc}
\braket{c^\dagger_{\mathbf{k}L}c_{-\mathbf{k}R}} = u^*v(\braket{b^\dagger_{-\mathbf{k}2} b_{-\mathbf{k}2}} - \braket{b^\dagger_{\mathbf{k}1} b_{\mathbf{k}1}})
\end{equation}
where, at finite temperature, the quasi-particles follow Fermi-Dirac statistics, that is,
\begin{eqnarray}\label{eq:bb}
\braket{b^\dagger_{\mathbf{k}1} b_{\mathbf{k}1}} = \frac{1}{e^{E^+/T}+1},\;\;\;\;
\braket{b^\dagger_{-\mathbf{k}2} b_{-\mathbf{k}2}} = \frac{1}{e^{E^-/T}+1}.
\end{eqnarray}
Note that both the BCS and CEH mechanisms for superconductivity are similar to those for neutrino or neutron--mirror neutron oscillations \cite{tan2019,tan2023a} in the sense of reaping the fruits of the misalignment between interaction and energy eigenstates.

Then, we can obtain the s-wave CEH gap equation from Eqs. \ref{eq:delta}, \ref{eq:ene}, \ref{eq:cc}, and \ref{eq:bb},
\begin{eqnarray}
\bm{\Delta} &=& \frac{V}{2}e^{-i\delta} \sum_\mathbf{k} \frac{\sinh(|\bm{\Delta}|/T)}{\cosh(|\bm{\Delta}|/T)+\cosh(\xi_\mathbf{k}/T)} \nonumber \\
&=& V\rho_F e^{-i\delta} \int^{\omega^*}_0 d\xi \frac{\sinh(|\bm{\Delta}|/T)}{\cosh(|\bm{\Delta}|/T)+\cosh(\xi/T)}
\end{eqnarray}
where $\rho_F$ denotes the density of states at the Fermi energy and the summation is replaced by an integration over the energy shell ($\pm \omega^*$) near the Fermi surface where the formation of superconducting pairs occurs. It is worth noting that $\omega^*$ bears resemblance to the Debye energy $\omega_D$ in the BCS theory. However, we will elaborate later on the more significant impact of $\omega^*$ within the CEH mechanism.

By working out the integration and introducing a dimensionless coupling parameter $\lambda = V\rho_F$ and a positive energy gap $\Delta$ defined by $\bm{\Delta}=\Delta e^{-i\delta}$, the s-wave gap equation can be simplified as,
\begin{eqnarray}
\Delta(T) &=& 2 \lambda T \tanh^{-1} \left( \tanh(\frac{\Delta(T)}{2T})\tanh(\frac{\omega^*}{2T}) \right) \nonumber \\
&=& \lambda T \log(\frac{e^{(\Delta(T)+\omega^*)/T}+1}{e^{\Delta(T)/T}+e^{\omega^*/T}}). \label{eq:sgap}
\end{eqnarray}

CEH Gap equations with more intricate orbital pairing symmetries can be calculated by considering an angular-dependent superconducting energy gap $\Delta_\mathbf{k}=\Delta \gamma_\mathbf{k}$. For a d-wave gap symmetry of $d_{x^2-y^2}$ in cuprate superconductors, we have the symmetry factor $\gamma_\mathbf{k} = \cos(2\varphi)$. A d-wave CEH gap equation can then be derived with ease,
\begin{equation}
\Delta(T) = \frac{8\lambda T}{\pi} \int^{\pi/4}_0 d\varphi \, \tanh^{-1}\left( \tanh(\frac{\Delta(T)\cos(2\varphi)}{2T})\tanh(\frac{\omega^*}{2T}) \right).
\label{eq:dgap}
\end{equation}

If we further consider $\bm{\Delta}$ as an emergent scalar field (akin to the Higgs field in particle physics), its self-interactions will lead to the same phenomenological Ginzburg-Landau theory as derived in BCS superconductivity.

\section{\label{gap}Analysis of CEH gap equations}

First, we should emphasize that both s-wave and d-wave gap equations in CEH (Eqs. \ref{eq:sgap} and \ref{eq:dgap}) are dramatically different from those derived in the BCS theory. In particular, the two parameters of $\omega^*$ and $\lambda$ play a crucial role in distinguishing the CEH mechanism from BCS.

To ensure that quasi-particles have positive energies (or negative energies for corresponding ``anti-particles'') as in Eq. \ref{eq:ene}, we establish the following superconducting requirement under the CEH model,
\begin{equation}\label{eq:sccondition}
\omega^* \leq \Delta(T)
\end{equation}
which is entirely different from that of BCS. In BCS, positive energies are guaranteed as $E^{+}_{\mathbf{k}} = \sqrt{\xi^2_{\mathbf{k}} + |\Delta|^2}$, and thus no constraint on the Debye energy $\omega_D$ is necessary. In addition, it should be noted that the coupling parameter $\lambda$ in CEH must be very large (i.e., $\lambda>1$ for s-wave and $\lambda>\pi/2$ for d-wave as presented below), as opposed to the small parameter of $\lambda \ll 1$ used in BCS. This implies that CEH is naturally suited for modeling superconductivity in strongly-correlated electron systems while BCS is more appropriate for the weak-coupling limit.

The condition of Eq. \ref{eq:sccondition} suggests that typically $\Delta(T_c)=\omega^*\neq 0$, meaning that the superconducting gap does not necessarily close at the critical temperature $T_c$, which is distinct from BCS. More details are presented below for both s- and d-wave cases.

\subsection{\label{sgap}s-wave results}

The CEH s-wave gap equation (Eq. \ref{eq:sgap}) can also be written as,
\begin{equation}
x^{1/\lambda}(x+w) = wx+1
\end{equation}
where $x=\exp(\Delta(T)/T)>1$ and $w=\exp(\omega^*/T)>1$. The superconducting condition of Eq. \ref{eq:sccondition} requires that $x>w$. We can then easily solve it for $\lambda$,
\begin{equation}
\lambda = \frac{\log(x)}{\log(\frac{wx+1}{w+x})}=\frac{\log(x)}{\log(x)+\log(\frac{w+1/x}{w+x})}>1
\end{equation}
which means that CEH addresses a strongly-correlated system.

\begin{figure}
\includegraphics[scale=1.0]{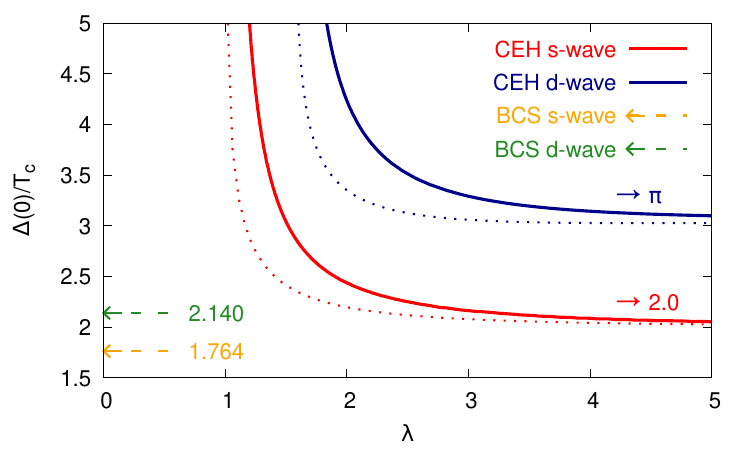}
\caption{\label{fig:d0tc} $\Delta_0/T_c$ as a function of $\lambda$ is shown for both s-wave and d-wave CEH superconductivity. BCS values in the weak-coupling limit are also displayed for comparison. The two dotted lines depict $\Delta_0/T_0$ corresponding to their respective CEH results.}
\end{figure}

In the limit of $T\rightarrow 0$, we obtain the gap at zero temperature,
\begin{equation}\label{eq:sd0}
\Delta_0\equiv \Delta(T=0) = \lambda \omega^*
\end{equation}
and the gap equation can then be simplified at the critical temperature $T_c$ as,
\begin{equation}
2\exp(\frac{\lambda+1}{\lambda^2}\frac{\Delta_0}{T_c})=\exp(\frac{2}{\lambda}\frac{\Delta_0}{T_c})+1.
\end{equation}
This allows us to plot the solution of $\Delta_0/T_c$ as a function of $\lambda$ as shown in Fig. \ref{fig:d0tc} with its two asymptotic limits: two as $\lambda\rightarrow \infty$ and $\log(2)\lambda^2/(\lambda-1)$ as $\lambda\rightarrow 1$. Note that this ratio is always larger than two and can become much larger at smaller $\lambda$ (or lower doping levels), in contrast to the value of about 1.764 in BCS for s-wave superconductors. To be clear, the ratios shown as arrows in the plot represent only the weak-coupling limit in BCS and higher $\Delta_0/T_c$ ratios are possible in extended models using stronger electron-phonon couplings. For simplicity, further comparisons to BCS in this paper will also be restricted to its weak-coupling limit unless stated otherwise.

\begin{figure}
\includegraphics[scale=1.0]{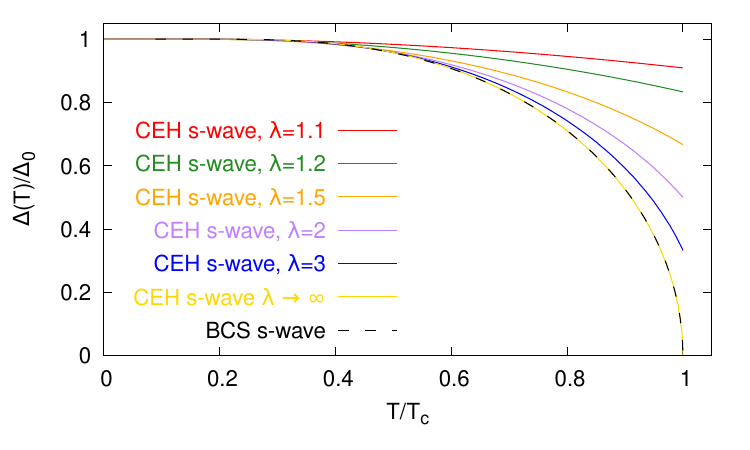}
\caption{\label{fig:swave} The $\Delta(T)/\Delta_0$ ratios as a function of $T/T_c$ are shown for s-wave CEH superconductivity. The gap only vanishes at $T_c$ when $\lambda\rightarrow \infty$, in which case the curve nearly overlaps with that of BCS.}
\end{figure}

Fig. \ref{fig:swave} shows the $\Delta(T)/\Delta_0$ ratios as a function of $T/T_c$ for various $\lambda$ values, which are numerically calculated from the gap equation. In Fig. \ref{fig:swave}, it is evident that the superconducting gap in CEH, in general, does not close at the critical temperature $T_c$, resulting in $\Delta(T_c)=\Delta_0/\lambda$ according to Eq. \ref{eq:sd0}. It only closes in the extreme case of $\lambda\rightarrow \infty$, where the relation between $\Delta(T)/\Delta_0$ and $T/T_c$ can be simplified as,
\begin{equation}
\frac{T}{T_c} = \frac{\Delta(T)/\Delta_0}{\tanh^{-1}(\Delta(T)/\Delta_0)}
\end{equation}
which nearly overlaps with the results from BCS as shown in Fig. \ref{fig:swave}. It is worth pointing out that, in this case, BCS extensions using strong electron-phonon couplings give very similar results to those obtained at its weak-coupling limit \cite{marsiglio2020}.

On the other hand, the superconducting gap does close at a higher temperature $T_0>T_c$. It can be calculated from the gap equation using the condition of $\Delta(T_0)=0$,
\begin{equation}
T_0 = \frac{\omega^*}{2\tanh^{-1}(1/\lambda)}=\frac{\omega^*}{\log(\frac{\lambda+1}{\lambda-1})}
\end{equation}
which, though higher, follows a similar trend as $T_c$ as demonstrated in the following section. Another ratio can be calculated simply as follows,
\begin{equation}
\Delta_0/T_0 = 2\lambda \tanh^{-1}(1/\lambda) < \Delta_0/T_c
\end{equation}
which is shown as the red dotted line in Fig. \ref{fig:d0tc}.

At $T\rightarrow 0$, we find, directly from the gap equation, that the energy gap exponentially approaches its maximum value of $\Delta_0$,
\begin{equation}
\Delta(T) \simeq \Delta_0 - \lambda T \exp(-\frac{\Delta_0}{T}(1-\frac{1}{\lambda}))
\end{equation}
which is similar to that of BCS.

\subsection{\label{dgap}d-wave results}

Taking the limit of $T\rightarrow 0$, we obtain the coupling parameter from the gap equation (Eq. \ref{eq:dgap}) (see Appendix \ref{app1}),
\begin{equation}\label{eq:dlam}
\lambda = \frac{\pi}{2(1-\sin2\theta)+4\theta \cos2\theta} > \frac{\pi}{2}
\end{equation}
where $\theta$ is defined by $\cos2\theta=\omega^*/\Delta_0$ within the range of $0<\theta<\pi/4$. This indicates that CEH d-wave superconductors require even stronger correlations. 

\begin{figure}
\includegraphics[scale=1.0]{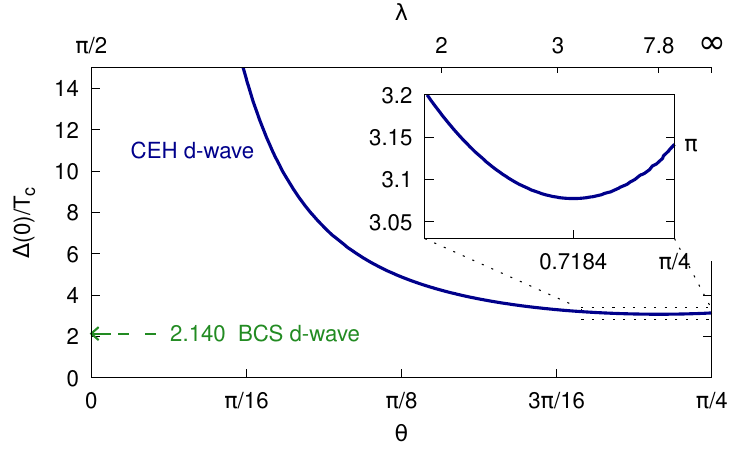}
\caption{\label{fig:d0tc2} The $\Delta_0/T_c$ ratios as a function of $\theta$ or $\lambda$ are shown for d-wave CEH superconductivity. The inset presents a minimum of $\approx 3.0774$ at $\theta\approx 0.7184$. The BCS d-wave value in the weak-coupling limit is also displayed for comparison.}
\end{figure}

We can also obtain the d-wave ratio of $\Delta_0/T_c$ numerically from the gap equation using $\omega^*=\Delta(T_c)=\Delta_0\cos2\theta$ and compare it with the s-wave and BCS results, as shown in Fig. \ref{fig:d0tc} and also in Fig. \ref{fig:d0tc2} as a function of $\theta$. The general trend of the d-wave ratio is similar to the s-wave one, though notably higher. However, the inset plot in Fig. \ref{fig:d0tc2} reveals that the d-wave ratio is not monotonic and has a minimum of about 3.0774 at $\theta\approx 0.7184$.

\begin{figure}
\includegraphics[scale=1.0]{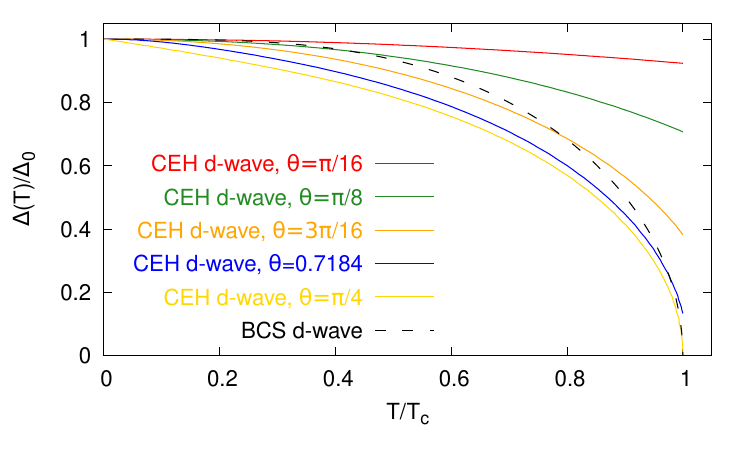}
\caption{\label{fig:dwave} The ratios of $\Delta(T)/\Delta_0$ as a function of $T/T_c$ are shown for d-wave CEH superconductivity. The gap only closes at $T_c$ when $\theta\rightarrow \pi/4$ or $\lambda\rightarrow \infty$. However, the gap-vanishing BCS curve behaves differently.}
\end{figure}

Similar to the s-wave results, the normalized d-wave superconducting gap as a function of $T/T_c$ is shown in Fig. \ref{fig:dwave}. Again, it does not close at $T_c$ which is contrary to the BCS prediction. Note that the temperature dependence of the BCS d-wave gap, unlike the s-wave case, differs greatly from the CEH predictions, even in the limit of $\lambda\rightarrow \infty$ or $\theta=\pi/4$. As a matter of fact, the distinction is so significant between BCS and CEH that measurements with decent experimental precision should be pursued.

It is straightforward to calculate the temperature $T_0>T_c$ where the d-wave superconducting gap vanishes,
\begin{equation}
T_0 = \frac{\omega^*}{2\tanh^{-1}(\pi/(2\lambda))}
\end{equation}
which is similar to the s-wave result except for a larger lower bound on the coupling parameter ($\lambda>\pi/2$). A similar ratio can also be obtained,
\begin{equation}
\Delta_0/T_0 = 2\tanh^{-1}(1+2\theta \cos 2\theta - \sin 2\theta)/\cos 2\theta < \Delta_0/T_c
\end{equation}
which is plotted as the blue dotted line in Fig. \ref{fig:d0tc}. Like $\Delta_0/T_c$, this ratio has a minimum of about 3.027 at $\theta\approx 0.662$.

The asymptotic behavior of the superconducting gap at $T\rightarrow 0$ can be obtained as follows,
\begin{equation}\label{eq:dddt}
\Delta \simeq \Delta_0 -\frac{T^2}{\theta\sin(2\theta)\cos(2\theta)\Delta_0}
\end{equation}
which is derived by utilizing the following integration,
\begin{equation}
 \left.\int^{\pi/4}_0 d\varphi \exp(-\frac{\Delta}{T}|\cos(2\varphi)-cos(2\theta)|)\right\vert_ {T\rightarrow 0} = \frac{T}{\Delta\sin(2\theta)}.
\end{equation}

\section{\label{sh}Entropy and Specific Heat}

In the CEH superconducting phase, the entropy of the finite-temperature system can be expressed through the statistics of Bogoliubov quasi-particles,
\begin{eqnarray}
S &=& -2\sum_\mathbf{k} (f_+\log f_+ + f_-\log f_-) \nonumber \\
 &=& -2\sum_\mathbf{k} (f_+\log f_+ + (1-f_+)\log(1-f_+))
\end{eqnarray}
where $f_{\pm}$ represent the Fermi-Dirac distributions of the quasi-particles as in Eq. \ref{eq:bb}. By replacing the summation with an energy integration, we obtain for the simple s-wave case,
\begin{equation}\label{eq:entropy}
S=2\int^{\Delta+\omega^*}_{\Delta-\omega^*} d\epsilon \rho(\epsilon) \left( \frac{\epsilon/T}{1+e^{\epsilon/T}}+\log(1+e^{-\epsilon/T}) \right)
\end{equation}
where $\rho(\epsilon)$ is the quasi-particle density of states. This entropy formula seems to be the same as the BCS one but only formally. The critical differences lie in $\rho(\epsilon)$ and the bounds of integration $\pm{\omega^*}$. In CEH, $\rho(\epsilon)=\rho_F$, whereas in BCS, the density exhibits a singular behavior at the gap energy, reflecting their differences in the pairing mechanism. More significantly, the integration bounds in CEH demand more careful handling, unlike BCS, due to a drastically different dispersion relation in Eq. \ref{eq:ene}. Such differences are most effectively showcased in the following calculations of heat capacity.

\subsection{\label{ssh}s-wave specific heat}

The specific heat for the CEH s-wave superconductors can be obtained from the entropy in Eq. \ref{eq:entropy} as follows,
\begin{equation}
C_{sc}=T\frac{\partial S}{\partial T}=2T\rho_F\int^{\Delta+\omega^*}_{\Delta-\omega^*} d\epsilon \frac{e^{\epsilon/T}}{(e^{\epsilon/T}+1)^2} \left(\left(\frac{\epsilon}{T}\right)^2 - \frac{\epsilon}{T}\frac{\partial \Delta(T)}{\partial T} \right)
\end{equation}
which can be simplified as,
\begin{equation}\label{eq:cswave}
C_{sc}(T)=2T\rho_F (s_2(T)-\frac{\partial \Delta(T)}{\partial T}s_1(T))
\end{equation}
where the two auxiliary functions are defined as
\begin{equation}
s_{1,2}(T) = \int^{(\Delta +\omega^*)/T}_{(\Delta -\omega^*)/T} dx \frac{e^x}{(e^x+1)^2}x^{1,2}.
\end{equation}

\begin{figure}
\includegraphics[scale=1.0]{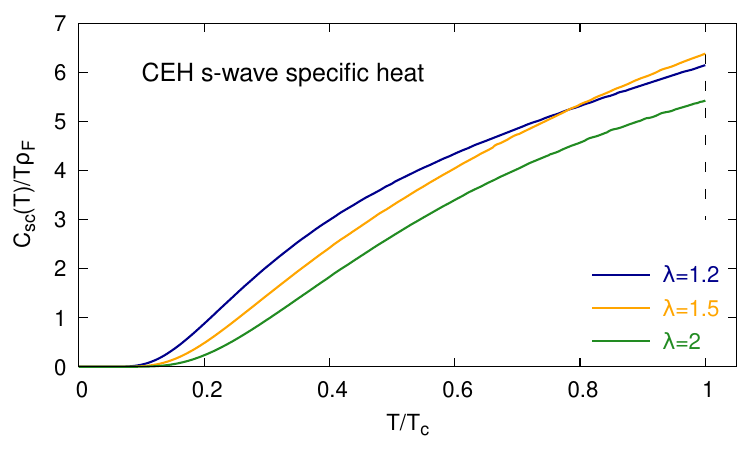}
\caption{\label{fig:cswave} The temperature dependence of heat capacity $C_{sc}(T)/(T\rho_F)$ is shown for s-wave CEH superconductivity. Cases with three different coupling parameters $\lambda=1.2,1.5,2$ are presented.}
\end{figure}

The CEH s-wave specific heat for various $\lambda$ values is shown in Fig. \ref{fig:cswave}, where it exponentially approaches zero in the limit of $T\rightarrow 0$,
\begin{equation}
C_{sc}(T) \rightarrow 2T\rho_F \left( \frac{\Delta_0-\omega^*}{T}\right)^2 e^{-(\Delta_0-\omega^*)/T}
\end{equation} 
and behaves similarly to the s-wave BCS superconductivity.

\begin{figure}
\includegraphics[scale=1.0]{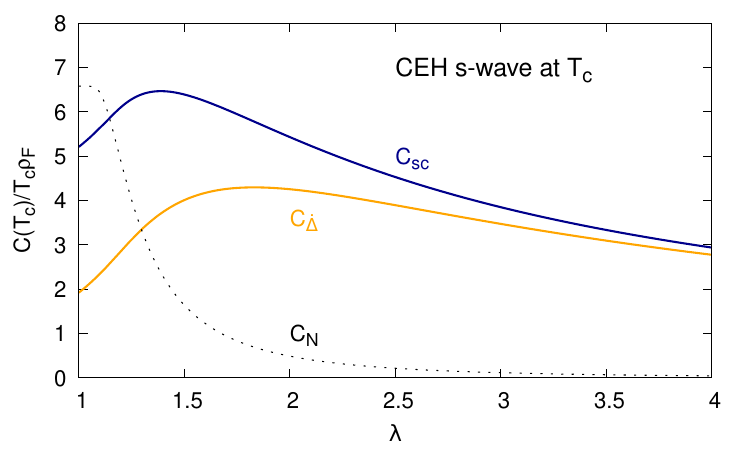}
\caption{\label{fig:shcswave} The heat capacity $C_{sc}(T_c)/(T_c\rho_F)$ at $T=T_c$ as a function of $\lambda$ is shown for s-wave CEH superconductivity. For comparison, the partial contribution $C_{\dot{\Delta}} $ from the second term in Eq. \ref{eq:cswave} and the normal-state heat capacity $C_N$ are also presented.}
\end{figure}

To better understand the heat capacity jump at the critical temperature, we plot its peak value at $T_c$ as a function of the coupling parameter $\lambda$ in Fig. \ref{fig:shcswave}. It is evident that the second term contribution $C_{\dot{\Delta}}$ in Eq. \ref{eq:cswave} dominates the heat capacity at larger $\lambda$. The normal state electronic heat capacity, within the same energy shell $\pm{\omega^*}$, can be written as,
\begin{equation}
C_N(T) = 2T\rho_F\int^{\omega^*/T}_{-\omega^*/T} dx \frac{e^{x}}{(e^{x}+1)^2} x^2
\end{equation}
where, in the bounds of integration, $\omega^*/T=\Delta_0/(\lambda T)$ decreases rapidly as $\lambda$ increases. The behavior of $C_N(T_c)$, as shown in Fig. \ref{fig:shcswave}, resembles that of normal metals only at very low $\lambda$ values (still $>1$), where $\omega^*/T_c$ remains large. It then drops rapidly to zero because $\omega^*/T_c$ approaches $2/\lambda$ asymptotically as $\lambda\rightarrow \infty$. Note that, unlike in the case of BCS, the first term in Eq. \ref{eq:cswave} does not reduce to $C_N$ at $T_c$ since the gap does not close at $T_c$ or $\Delta(T_c)\neq 0$.

Fig. \ref{fig:shcswave} demonstrates that optimal superconductivity for s-wave pairing favors relatively weaker correlations ($\lambda \lesssim 1.5$). The specific capacity jump at $T_c$ from the superconducting state to the normal state is more complicated in CEH. The normal state electronic heat capacity in BCS maintains a constant $C_N/T$, giving rise to a fixed jump ratio of about 1.43 for s-wave pairing. However, in CEH, this ratio varies depending on $\lambda$ or the doping level and may reach about 3 at $\lambda \sim 1.5$ as $C_N$ decreases. More complicatedly, the $\omega^*$ energy shell may shrink so significantly at larger $\lambda$ or higher doping levels that other energy bands could become accessible for electrons, causing an increase in $C_N$ and resulting in a smaller heat capacity jump at $T_c$.

\subsection{\label{dsh}d-wave specific heat}

The d-wave specific heat can be similarly obtained as follows,
\begin{equation}\label{eq:shd}
C_{sc}(T)=\frac{8T\rho_F}{\pi} \left(  \int^{\pi/4}_0 d\varphi d_2(\varphi,T)-\frac{\partial \Delta}{\partial T} \int^{\pi/4}_0 d\varphi d_1(\varphi,T)\cos(2\varphi) \right) 
\end{equation}
where the two integrand functions, similar to the s-wave case, are defined by
\begin{equation}
d_{1,2}(\varphi,T) = \int^{(\Delta \cos(2\varphi)+\omega^*)/T}_{(\Delta \cos(2\varphi)-\omega^*)/T} dx \frac{e^x}{(e^x+1)^2}x^{1,2}
\end{equation}
and the derivative of the gap can be obtained by differentiating the gap equation,
\begin{equation}
\frac{\partial \Delta}{\partial T} = \frac{\Delta}{T}+\frac{\omega^*}{T} \frac{4\lambda g_2(T)}{\pi\exp(-\omega^*/T)-8\lambda \sinh(\omega^*/T) g_1(T)}
\end{equation}
where the two auxiliary functions are given by
\begin{eqnarray}
g_1(T)&=&\int^{\pi/4}_0 d\varphi \frac{e^{\Delta\cos(2\varphi)/T}\cos(2\varphi)}{(e^{(\Delta\cos(2\varphi)+\omega^*)/T}+1)(e^{\Delta\cos(2\varphi)/T}+e^{\omega^*/T})} \nonumber \\
g_2(T)&=&\int^{\pi/4}_0 d\varphi \frac{1-e^{2\Delta\cos(2\varphi)/T}}{(e^{(\Delta\cos(2\varphi)+\omega^*)/T}+1)(e^{\Delta\cos(2\varphi)/T}+e^{\omega^*/T})}.
\end{eqnarray}

\begin{figure}
\includegraphics[scale=1.0]{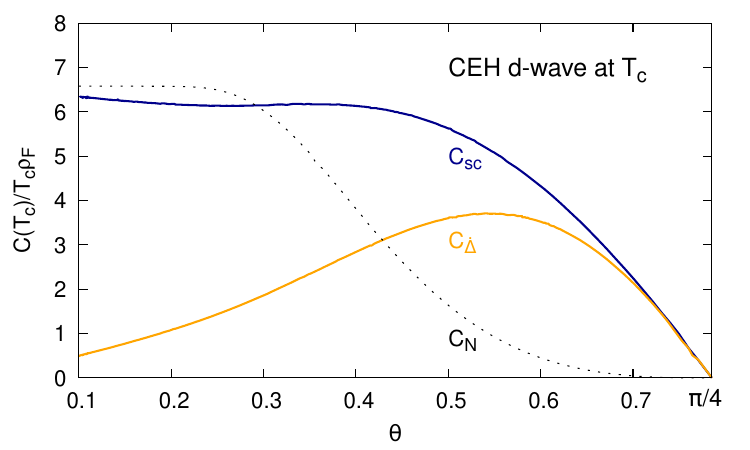}
\caption{\label{fig:shcdwave} The heat capacity $C_{sc}(T_c)/(T_c\rho_F)$ at $T=T_c$ as a function of $\theta$ is shown for d-wave CEH superconductivity. For comparison, the partial contribution $C_{\dot{\Delta}} $ from the second term in Eq. \ref{eq:shd} and the normal-state electronic heat capacity $C_N$ are also presented.}
\end{figure}

As presented in Fig. \ref{fig:shcdwave}, the peak specific heat at $T=T_c$ for d-wave superconductivity follows the trend of s-wave, though large heat capacity jumps are extended to larger $\theta$ / $\lambda$ or higher doping levels. Again, compared to a constant jump ratio of about 0.95 for BCS d-wave, the normalized jump ratio in CEH varies (reaching $\sim 1.2$ at $\lambda=2$). The d-wave ratio is generally lower than the s-wave one due to a slower decrease in $C_N$ as $\theta$ and $\lambda$ increase.

\begin{figure}
\includegraphics[scale=1.0]{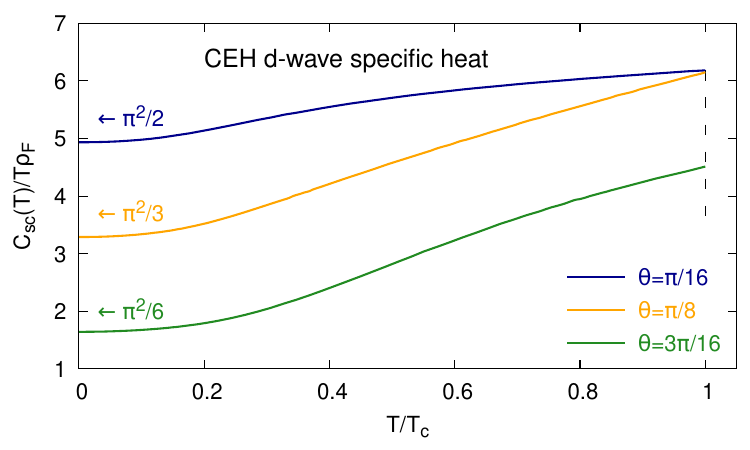}
\caption{\label{fig:cdwave} The temperature dependence of heat capacity $C_{sc}(T)/(T\rho_F)$ is shown for d-wave CEH superconductivity. Cases with three different coupling strengths $\theta=\pi/16,\pi/8,3\pi/16$ are presented.}
\end{figure}

One notable finding in CEH is its unique prediction of a non-zero linear term in the d-wave heat capacity at the zero-temperature limit, as shown in Fig. \ref{fig:cdwave}. This non-zero offset originates from the $d_2(\varphi,T)$ term of Eq. \ref{eq:shd} (see Appendix \ref{app2}),
\begin{equation}\label{eq:cd0}
\gamma(0)\equiv\left.\dfrac{C_{sc}(T)}{T}\right\vert_{T\rightarrow 0} = \dfrac{8\pi}{3}\left( \dfrac{\pi}{4}-\theta\right)\rho_F \neq 0
\end{equation}
which explains the ``anomalous linear term'' observed in the heat capacity of cuprates \cite{ramirez1987,kato1988}. It may not approach zero even at the extreme overdoping limit where $\theta=\pi/4$ or $\lambda\rightarrow \infty$. Assuming a constant four-fermion interaction strength, i.e., $\lambda \varpropto \rho_F$, and taking into account Eq. \ref{eq:dlam}, we see that $\gamma(0)$ decreases as the doping level or $\lambda$ increases, and we obtain its underdoping-to-overdoping ratio $\gamma(0)_{UL}/\gamma(0)_{OL} = \pi^2/4 \approx 2.47$ calculated at the two extreme doping limits, which agrees very well with the cuprate measurements \cite{kato1988, wen2004}.

The next leading order at $T\rightarrow 0$ in Fig. \ref{fig:cdwave} is contributed from both terms (see Appendix \ref{app2}),
\begin{equation}\label{eq:cd2}
\dfrac{C_{sc}(T)}{T\rho_F}-\dfrac{\gamma(0)}{\rho_F} \simeq \left(\dfrac{8\pi}{3\theta} + \dfrac{14\pi^3}{15\tan2\theta} \right) \dfrac{1}{\sin^22\theta}\left( \dfrac{T}{\Delta_0}\right)^2 
\end{equation}
where the first part within the parentheses is calculated from the $d_1(\varphi,T)$ term and the second from the $d_2(\varphi,T)$ term. This exactly explains the quadratic behavior observed in the heat capacity curve of $C/T$ at $T\ll T_c$ in d-wave superconducting cuprates \cite{loram1994,moler1994,revaz1998}.

\section{\label{exp}Discussions and Comparisons with Experimental Data}

The proposed CEH mechanism provides many unique predictions that are distinct from the well known BCS results. These differences have been discussed above in detail and are also summarized in Table \ref{tab:1}. In particular, CEH naturally explains the necessity of antiferromagnetism and strong coupling in non-BCS superconductivity. Most strikingly, many of its predictions agree very well with existing experimental data on non-BCS superconductors, particularly cuprates and FeSCs. Several such examples will be illustrated below.

\begin{table*}
\caption{\label{tab:1} Summary of comparisons of some major results between BCS and CEH. The results from BCS are calculated in the weak-coupling limit and other extensions using stronger electron-phonon couplings may differ.}
\begin{ruledtabular}
\begin{tabular}{c|cccc}
Model & \multicolumn{2}{c}{BCS} & \multicolumn{2}{c}{CEH} \\
\hline
material & \multicolumn{2}{c}{averse to magnetism} & \multicolumn{2}{c}{desirous of antiferromagnetism} \\
mechanism & \multicolumn{2}{c}{Cooper pairs} & \multicolumn{2}{c}{chiral electron-hole pairs} \\
SC criteria & \multicolumn{2}{c}{no constraint on Debye $\omega_D$} & \multicolumn{2}{c}{$\omega^* < \Delta(T)$} \\
\hline
symmetry & s-wave & d-wave & s-wave & d-wave \\
\cline{2-5}
\multirow{2}{*}{coupling} & \multicolumn{2}{c}{weak coupling} & \multicolumn{2}{c}{strong coupling} \\
& \multicolumn{2}{c}{$\lambda \ll 1$} & $\lambda > 1$ & $\lambda > \frac{\pi}{2}$, $0<\theta<\frac{\pi}{4}$ \\
\cline{2-5}
$E^{\pm}_{\mathbf{k}}=$ &  $\pm\sqrt{\xi^2_{\mathbf{k}}+|\Delta|^2}$ &  $\pm\sqrt{\xi^2_{\mathbf{k}}+|\Delta\cos2\varphi|^2}$ & $\xi_{\mathbf{k}} \pm |\Delta|$ & $\xi_{\mathbf{k}} \pm |\Delta\cos2\varphi|$ \\
\multirow{2}{*}{gap equation} & \multirow{2}{*}{$\frac{1}{\lambda}=\int^{\omega_D}_0 d\xi \frac{\tanh(E^+/2T)}{E^+}$} & $\frac{1}{\lambda}=\int^{2\pi}_0 d\varphi \frac{\cos^22\varphi}{2\pi}\times$ & \multirow{2}{*}{Eq. \ref{eq:sgap}} & \multirow{2}{*}{Eq. \ref{eq:dgap}} \\
& & $\int^{\omega_D}_0 d\xi \frac{\tanh(E^+/2T)}{E^+}$ & & \\
\cline{2-5}
$\Delta_0=$ & $2\omega_D e^{-1/\lambda}$ & $2.426\omega_D e^{-2/\lambda}$ & $\lambda\omega^*$ & $\omega^*/\cos2\theta$ \\
\multirow{2}{*}{$T_c$ or $T_0=$} & \multicolumn{2}{c}{$T_c=T_0$} & \multicolumn{2}{c}{$T_c<T_0$} \\
	& $T_c=1.134\omega_D e^{-1/\lambda}$ & $T_c=1.134\omega_D e^{-2/\lambda}$ & $T_0=\frac{\omega^*}{2\tanh^{-1}(1/\lambda)}$ & $T_0=\frac{\omega^*}{2\tanh^{-1}(\pi/2\lambda)}$ \\
$\Delta_0/T_c=$ & 1.764 & 2.140 & $>2$ (Fig. \ref{fig:d0tc}) & $\gtrsim 3.077$ (Fig. \ref{fig:d0tc2}) \\
$\Delta(T_c)$ & \multicolumn{2}{c}{$\Delta(T_c)=0$, gap closes at $T_c$} & \multicolumn{2}{c}{$\Delta(T_c)=\omega^* \neq0$, does not close at $T_c$} \\
$\frac{\Delta(T)}{\Delta_0}$ vs. $\frac{T}{T_c}$ & dashed line in Fig. \ref{fig:swave} & dashed line in Fig. \ref{fig:dwave} & Fig. \ref{fig:swave} & Fig. \ref{fig:dwave} \\
\cline{2-5}
specific heat & & & Figs. \ref{fig:shcswave}-\ref{fig:cswave} & Figs. \ref{fig:shcdwave}-\ref{fig:cdwave} \\
\multirow{2}{*}{$C/T(T\rightarrow 0)$} & exponentially& linearly & exponentially & quadratically \\
& approaches 0 & approaches 0 & approaches 0 & to $\frac{8\pi}{3}(\frac{\pi}{4}-\theta)\rho_F\neq 0$ \\
jump at $T_c$ & 1.43 & 0.95 & varies with $\lambda$ & varies with $\theta$ \\
\end{tabular}
\end{ruledtabular}
\end{table*}

CEH conducts current via relative motion and this may explain why flat bands are favored in high-$T_c$ superconductivity. The key feature in CEH is the $\omega^*$ energy shell, which could function as both the superconducting band for CEH pairs and an energy gap (closely related to the widely-recognized pseudogap) for normal state electrons. This $\omega^*$ band may be one of the flat bands where center-of-mass motion is forbidden, making it ideal for the formation of antiferromagnetism and CEH pairs. For easier comparison with data below, we assume a simple linear relationship between $\omega^*$ and the pseudogap temperature $T^*$: $\omega^* \varpropto T^* $.

In the undoped parent compound, $\omega^*$ typically exceeds the superconducting gap $\Delta$, which can disrupt the stability of CEH pairs with additional energy. Doping, however, plays a crucial role in reducing $\omega^*$, making it below the level of $\Delta$, and thereby facilitating the onset of superconductivity. Meanwhile, doping tends to increase the coupling parameter $\lambda$ as the density of states in the $\omega^*$ band rises due to unitarity or conservation of the number of quantum states in a compressed $\omega^*$ band. High-pressure-induced superconductivity, investigated in various materials, may introduce similar effects by compressing the $\omega^*$ band with external pressure.

\begin{figure}
\includegraphics[scale=1.0]{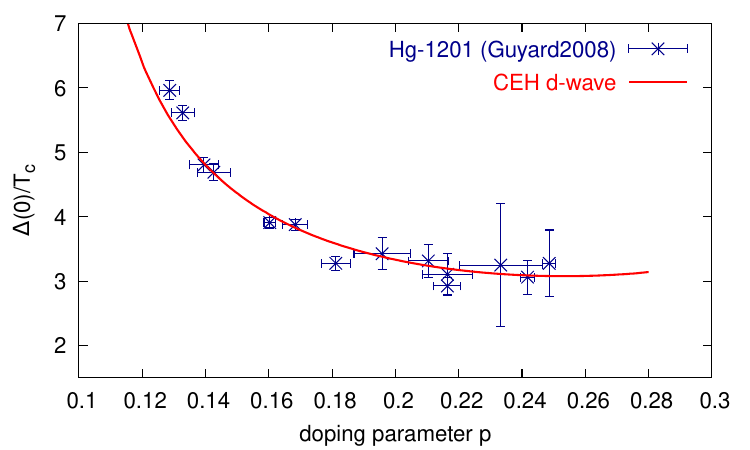}
\caption{\label{fig:hg1201} $\Delta_0/T_c$ as a function of the doping level predicted by d-wave CEH is shown in good agreement with the HgBa$_2$CuO$_{4+\delta}$ (Hg-1201) data \cite{guyard2008}, which are much higher than the BCS weak-coupling limit value of 2.140.}
\end{figure}

It have been observed in various cuprate superconductors that the ratio $\Delta_0/T_c$ exceeds three. This ratio has been shown to undergo a dramatic increase with decreasing doping and approach a limit of three near maximum doping (see Fig. 3 of Ref. \cite{hawthorn2007} and the references therein). This behavior aligns well with our d-wave prediction based on the new pairing mechanism as shown in Fig. \ref{fig:d0tc}. To further demonstrate this, a direct comparison between CEH predictions and experimental data for HgBa$_2$CuO$_{4+\delta}$ (Hg-1201) \cite{guyard2008} is presented in Fig. \ref{fig:hg1201}. A good fit to the experimental data is achieved using a simple $\lambda-p$ parametrization discussed below. Furthermore, large $\Delta_0/T_c$ ratios, consistent with CEH s-wave predictions, have also been observed in iron-based superconductors \cite{zhang2012}.

It has been observed in various studies (e.g., \cite{suzuki1999,guyard2008,ren2012}) that the non-BCS superconducting gap does not close at the critical temperature, exhibiting a behavior similar to that shown in Fig. \ref{fig:dwave}. In particular, in Fig. 3 of Ref. \cite{guyard2008}, the trend of the temperature dependence closely resembles our results and also displays a similar deviation from the BCS prediction.

Another piece of clear and convincing evidence comes from the detailed analysis of d-wave specific heat presented above, which is worth reiterating. CEH predicts a non-zero linear term in heat capacity at $T=0$ for d-wave superconductors, a phenomenon well observed in many cuprates \cite{ramirez1987,kato1988, wen2004}. Furthermore, the predicted quadratic temperature dependence in $C/T$ near zero temperature is also in agreement with observations \cite{loram1994,moler1994,revaz1998}, in stark contrast to the BCS linear dependence.

To make direct comparisons between CEH predictions and the wealth of data accumulated over decades of high-$T_c$ superconductivity studies, it is necessary to establish a concrete relationship between the coupling parameter $\lambda$ and the doping level parameter $p$. We will start with some rough yet simple estimates to facilitate direct comparisons to the data. For strongly correlated parent compounds like cuprates or FeSCs, it is reasonable to assume that the initial coupling parameter $\lambda \sim 1$. Doping then increases the density of charge carrier states, effectively making $\lambda$ larger. It has been observed that the carrier density increases very rapidly at very low doping levels and then more gradually at higher doping levels \cite{cooper1990}. At low doping levels or in the case of CEH s-wave superconductors, we can approximate this with the following $\lambda-p$ relation
\begin{equation}\label{eq:lamp}
\lambda = 1-\frac{1}{\log(p/p_m)}
\end{equation}
where $p_m$ is the maximum doping level corresponding to $\lambda\rightarrow \infty$. For the example of an s-wave FeSC discussed below, we choose to adopt $p_m=1/3$.

However, CEH d-wave superconductors require larger correlations, specifically, $\lambda>\pi/2$, which means that superconductivity will not occur until the doping reaches a minimum level $p_0$. To describe d-wave superconductors, we can apply the following parametrization,
\begin{equation}\label{eq:thep}
p=p_0 + a \sin^2\theta.
\end{equation}
For the comparison shown in Fig. \ref{fig:hg1201}, parameters of $p_0=0.08$ and $a=0.4$ are adopted.

\begin{figure}
\includegraphics[scale=1.0]{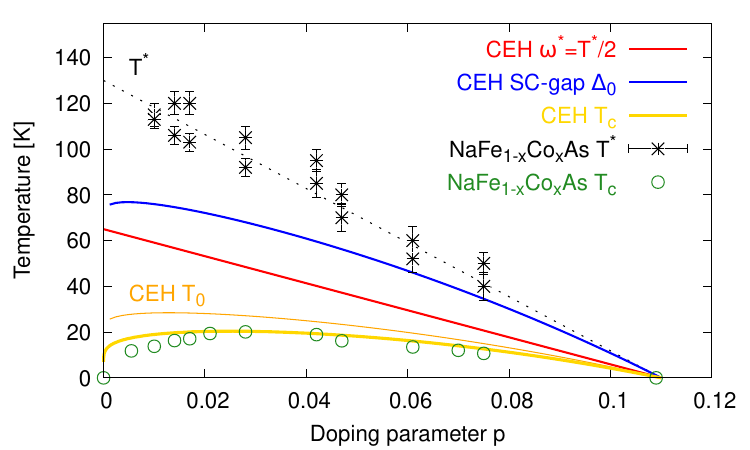}
\caption{\label{fig:fesc} Phase diagram, predicted by CEH assuming that $p_m=1/3$ and $\omega^*=0.5T^*$, is compared to experimental data on NaFe$_{1-x}$Co$_x$As \cite{wang2012, wang2013}.}
\end{figure}

Using the $\lambda-p$ and $\theta-p$ relations given in Eqs. \ref{eq:lamp}-\ref{eq:thep}, we can compare the phase diagrams constructed from CEH to experimental data in two examples. The first example involves NaFe$_{1-x}$Co$_x$As with experimental data taken from Refs. \cite{wang2012, wang2013}. Fig. \ref{fig:fesc} shows the results from CEH assuming that $p_m=1/3$ and $\omega^*=0.5T^*$. The superconducting phase and $T_c$ values derived from s-wave CEH agree well with the data. One possible issue is that $T_c$ appears moderately overestimated at extremely low doping levels ($p \lesssim 0.01$), which could stem from the oversimplified parametrization in Eq. \ref{eq:lamp}. The $\lambda-p$ relation may require more nuanced treatment at extremely low doping levels where the density of states undergoes the most significant changes. Other parameters such as the superconducting gap $\Delta_0$ and the gap closing temperature $T_0$ predicted by s-wave CEH are also depicted in Fig. \ref{fig:fesc}.

\begin{figure}
\includegraphics[scale=1.0]{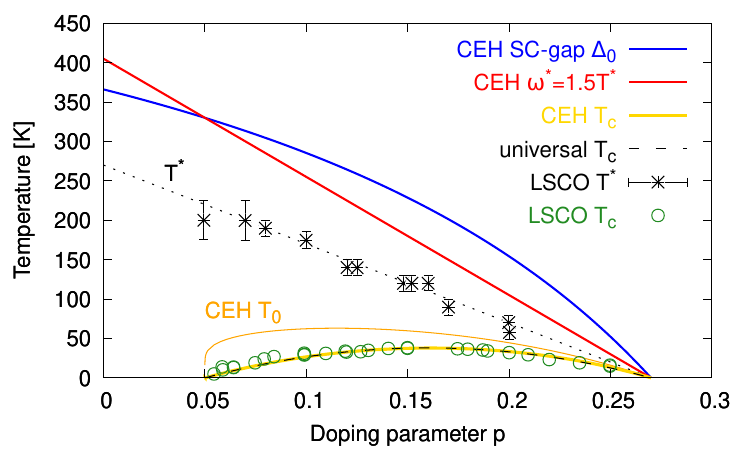}
\caption{\label{fig:cusc} Phase diagram, predicted by CEH assuming that $p=0.05+0.94 \sin^2\theta$ and $\omega^*=1.5T^*$, is compared to experimental data on La$_{2-x}$Sr$_x$CuO$_4$ (LSCO) \cite{yamada1998,doiron-leyraud2012}.}
\end{figure}

Another example is the phase diagram of La$_{2-x}$Sr$_x$CuO$_4$ (LSCO). In Fig. \ref{fig:cusc}, a comparison between the measured LSCO data \cite{yamada1998,doiron-leyraud2012} and the d-wave predictions from CEH ($p=0.05+0.94 \sin^2\theta$ and $\omega^*=1.5T^*$) is presented. The superconducting phase, characterized by the well-known dome shape for cuprates, is well reproduced and CEH shows good agreement with both the data \cite{yamada1998} and the universal $T_c$ parametrization of $T_c/T^{\textrm{max}}_c=1-82.6(p-0.16)^2$ \cite{presland1991} using $T^{\textrm{max}}_c=38 K$ for LSCO.

The same $\theta-p$ relation applied in LSCO, along with $\omega^*=2.3T^*$, also performs equally well in predicting the phase diagram of YBCO, despite significantly higher $T^*$ and $T_c$ values in YBCO. More systematic comparisons, especially to experimental data on $\Delta_0$ and $T_0$, will provide more compelling evidence for CEH. Further comparisons with other non-BCS superconducting materials would offer more valuable insights into understanding these parametrization relations.

Note that the $T^*$ lines shown in both Figs. \ref{fig:fesc} and \ref{fig:cusc} do not agree with an alternative view where the $T^*$ line could cross into the superconducting phase and end at a smaller critical doping point. This is because the pseudogap phase between the $T^*$ and $T_c$ lines is related to superconductivity, that is, the pseudogap is not an independent order competing with superconductivity in the CEH mechanism. It could possibly be understood as a short-range antiferromagnetic order in the phase between $T^*$ and $T_0$, e.g., a liquid of antiferromagnetic singlet dimer states as proposed in the resonating valence bond theory \cite{anderson1987}. When it enters the phase between $T_0$ and $T_c$, these antiferromagnetic dimers could transition into CEH dimers due to the appearance of superconducting gap. However, these preformed CEH pairs could not superconduct owing to the condition of $\omega^* > \Delta(T)$ in this phase, which could easily convert these pairs back to normal antiferromagnetic dimers. The long-range CEH order could only be established for unconventional superconductivity under the condition of $\omega^* < \Delta(T)$ at $T<T_c$.

\section{\label{outlook}Conclusions and Outlook}
Using the new chiral electron-hole pairing mechanism, we provide a more comprehensive understanding of non-BCS superconductivity in a strongly correlated electron system. Our new predictions are remarkably consistent with numerous puzzling properties observed in cuprate and FeSC superconductors such as the unexpectedly large $\Delta_0/T_c$ ratios, the absence of gap closure at $T_c$, the presence of a non-zero $\gamma(0)$ term and a quadratic trend in the heat capacity ratio of $C/T$ as $T\rightarrow 0$ in cuprates, among others. Further measurements (e.g., on $T_0$) and systematic comparisons with experimental data across diverse material types will provide more stringent tests on the CEH mechanism. A better understanding of the $\omega^*$ band and the $\lambda$-doping level relationship may help identify even more promising high-$T_c$ superconducting materials in the near future.

\begin{acknowledgments}
This work is supported in part by the faculty research support program at the University of Notre Dame. The author would like to thank Edwin Huang for his insightful comments.
\end{acknowledgments}

\appendix
\section{d-wave $\lambda-\theta$ relation \label{app1}}
The d-wave gap equation (Eq. \ref{eq:dgap}) can be rewritten as,
\begin{equation}
\Delta(T) = \frac{4\lambda T}{\pi} \int^{\pi/4}_0 d\varphi \, \log \left( \frac{e^{(\Delta\cos 2\varphi+\omega^*)/T}+1}{e^{\Delta\cos 2\varphi/T}+e^{\omega^*/T}} \right).
\label{eq:dgap2}
\end{equation}
As $T\rightarrow 0$, the integrand becomes $\Delta\cos 2\varphi/T$ if $\Delta\cos 2\varphi<\omega^*$ and $\omega^*/T$ otherwise. By introducing a new parameter $\theta$ with 
\begin{equation}\label{eq:theta}
\cos(2\theta)=\omega^*/\Delta_0
\end{equation}
where the range of $\theta$ is limited to $0<\theta<\pi/4$, we obtain in the zero-temperature limit
\begin{eqnarray}
\Delta_0 &=& \frac{4\lambda T}{\pi} \left( \int^{\theta}_0 d\varphi \frac{\omega^*}{T} + \int^{\pi/4}_{\theta} d\varphi \frac{\Delta_0\cos 2\varphi}{T} \right) \nonumber \\
&=& \frac{4\theta}{\pi} \lambda \omega^* + \frac{2(1-\sin 2\theta)}{\pi}\lambda\Delta_0
\end{eqnarray}
which, by using Eq. \ref{eq:theta}, immediately leads to the d-wave $\lambda-\theta$ relation as given in Eq. \ref{eq:dlam}.

\section{d-wave specific heat at $T\rightarrow 0$ \label{app2}}
The d-wave specific heat in Eq. \ref{eq:shd} can be expressed as a sum of two terms: $C_{sc} = C_2 +C_1$ where $C_{2,1}$ involve the integration of the two functions $d_{2,1}(\varphi,T)$, respectively. For both terms, the integration with respect to $\varphi$ can be divided into two parts: from 0 to $\theta'$ and from $\theta'$ to $\pi/4$, where $\cos 2\theta' = \omega^*/\Delta$. For convenience, we introduce the following definitions,
\begin{eqnarray}
y_{\pm} &\equiv& (\cos 2\varphi \pm \cos 2\theta')\Delta/T \\
h_n(x) &\equiv& \frac{e^x}{(e^x+1)^2}x^n.
\end{eqnarray}

\subsection{$C_1$ contribution at $T\rightarrow 0$}

First, we work with the $C_1$ term where we can simplify the second part of the integral as follows,
\begin{eqnarray}
\int^{\pi/4}_{\theta'} d\varphi \cos 2\varphi \int^{y_+}_{y_-}dx h_1(x) &=& \int^{\pi/4}_{\theta'} d\varphi \cos 2\varphi \left( \int^{y_+}_{-y_+} - \int^{-|y_-|}_{-y_+} \right)dx h_1(x) \nonumber \\
&=& \int^{\pi/4}_{\theta'} d\varphi \cos 2\varphi \int^{y_+}_{|y_-|} dx h_1(x)
\end{eqnarray}
because $h_1(x)$ is an odd function. By taking the limit of $y_+\rightarrow \infty$ as $T\rightarrow 0$ and the following integration,
\begin{equation}
\int^{\infty}_y dx h_1(x) = \log(1+e^y)-ye^y/(e^y+1) \equiv j(y),
\end{equation}
where $j(y)$ is odd as well, we can obtain the full integration as,
\begin{eqnarray}
\int^{\pi/4}_{0} d\varphi \cos 2\varphi \int^{y_+}_{y_-}dx h_1(x) &=& \left(\int^{\theta'}_{0} \int^{y_+}_{y_-} + \int^{\pi/4}_{\theta'} \int^{y_+}_{|y_-|} \right) d\varphi dx \cos 2\varphi h_1(x) \nonumber \\
&=& \left(\int^{\theta'}_{0} + \int^{\pi/4}_{\theta'} \right) d\varphi \cos 2\varphi j(|y_-|).
\end{eqnarray}
By a change of variable from $\varphi$ to $y_-$, we obtain for the first part,
\begin{eqnarray}
\int^{\theta'}_{0} d\varphi \cos 2\varphi j(y_-) &=& \int^{(1-\cos 2\theta')\Delta/T}_0 dy_- j(y_-)\frac{T}{2\tan 2\varphi \Delta} \nonumber \\
&=& \frac{\pi^2T}{12\tan 2\theta' \Delta}
\end{eqnarray}
where the main contribution arises from the vicinity of $\varphi \sim \theta'$, allowing us to approximate $\tan 2\varphi$ with $\tan 2\theta'$, and the upper limit of integration becomes $\infty$ as $T\rightarrow 0$, allowing us to use the identity $\int^{\infty}_0 dy j(y) = \pi^2/6$. Likewise, the second part of the integral yields the same result.

From Eq. \ref{eq:dddt}, we can obtain the derivative of the gap,
\begin{equation}
\frac{\partial \Delta(T)}{\partial T} = -\frac{2T}{\theta'\sin(2\theta')\cos(2\theta')\Delta}.
\end{equation}

Putting them all together, we can obtain the $C_1$ contribution to the specific heat,
\begin{equation}
\frac{C_1}{T\rho_f} = \dfrac{8\pi}{3\theta' \sin^22\theta'} \left( \dfrac{T}{\Delta}\right)^2
\end{equation}
which gives the first term in Eq. \ref{eq:cd2} by replacing $\theta'$ and $\Delta$ with $\theta'=\theta$ and $\Delta = \Delta_0$ at $T=0$.

\subsection{$C_2$ contribution at $T\rightarrow 0$}

Similarly, contributions to the $C_2$ term can be separated into three parts,
\begin{eqnarray}\label{eq:c2term}
\int^{\pi/4}_{0} d\varphi \int^{y_+}_{y_-}dx h_2(x) &=& \left(\int^{\theta'}_{0} \int^{y_+}_{y_-} - \int^{\pi/4}_{\theta'} \int^{y_+}_{|y_-|} + \int^{\pi/4}_{\theta'} \int^{y_+}_{-y_+} \right) d\varphi dx h_2(x) \nonumber \\
&\equiv& C'_{2a} + C'_{2b} + C'_{2c}
\end{eqnarray}
because $h_2(x)$ is an even function.

We can easily integrate out the last part by taking $y_+ \rightarrow \infty$ as $T\rightarrow 0$ and $\int^{\infty}_{-\infty} dx h_2(x) = \pi^2/3$,
\begin{equation}
C'_{2c}=\int^{\pi/4}_{\theta'} d\varphi \int^{\infty}_{-\infty}dx h_2(x) = \frac{\pi^2}{3}(\frac{\pi}{4}-\theta')
\end{equation}
which gives the non-zero offset in Eq. \ref{eq:cd0}.

The first part of Eq. \ref{eq:c2term} can be worked out by changing the order of integration as follows,
\begin{equation}
C'_{2a} = \left( \int^{x_1}_0 \int^{Tx/\Delta+\cos 2\theta'}_{\cos 2\theta'} + \int^{x_2}_{x1} \int^{1}_{\cos 2\theta'} + \int^{x_3}_{x2} \int^{1}_{Tx/\Delta-\cos 2\theta'} \right) dx h_2(x) \frac{d\cos 2\varphi}{2\sin 2\varphi}
\end{equation}
where $x_1=(1-\cos 2\theta')\Delta/T$, $x_2=2\cos 2\theta' \Delta/T$, or vice versa, and $x_3=(1+\cos 2\theta')\Delta/T$. As $T\rightarrow 0$, we have $x_{1,2,3} \rightarrow \infty$, rendering the last two terms in the above integral negligible. The integration with respect to $\varphi$ in the first term results in $\theta' -\cos^{-1}(Tx/\Delta+\cos 2\theta')/2$. Using the following expansion,
\begin{equation}
\cos^{-1}(t+\cos 2\theta) = 2\theta - \frac{t}{\sin 2\theta} - \frac{\cos 2\theta t^2}{2\sin^3 2\theta} + O(t^3),
\end{equation}
we can simplify the integral as
\begin{equation}\label{eq:c21}
C'_{2a} = \frac{T}{2\sin 2\theta' \Delta} \int^{\infty}_0dx h_3(x) + \frac{\cos 2\theta'}{4\sin^3 2\theta}\left(\frac{T}{\Delta}\right)^2 \int^{\infty}_0dx h_4(x) + O(T^3).
\end{equation}

The second part of Eq. \ref{eq:c2term} can be treated similarly as,
\begin{equation}
C'_{2b} = -\left( \int^{x_4}_0 dx h_2(x) \int^{\cos 2\theta'}_{\cos 2\theta'-Tx/\Delta} + \int^{x_2}_{x_4} dx h_2(x) \int^{\cos 2\theta'}_{Tx/\Delta-\cos 2\theta'} \right) \frac{d\cos 2\varphi}{2\sin 2\varphi}
\end{equation}
where $x_4=\cos 2\theta' \Delta/T$ and again, only the first term contributes. Then it can be worked out as,
\begin{equation}\label{eq:c22}
C'_{2b} = -\frac{T}{2\sin 2\theta' \Delta} \int^{\infty}_0dx h_3(x) + \frac{\cos 2\theta'}{4\sin^3 2\theta}\left(\frac{T}{\Delta}\right)^2 \int^{\infty}_0dx h_4(x) + O(T^3).
\end{equation}

The linear terms in Eqs. \ref{eq:c21} and \ref{eq:c22} exactly cancel each other out. By summing up the two parts and using $\int^{\infty}_0dx h_4(x) = 7\pi^4/30$, we obtain the $C_2$ contribution to the quadratic temperature dependence, which gives the second term in Eq. \ref{eq:cd2}.

\bibliography{sc}

\end{document}